\documentclass[10.5pt]{article}
\usepackage{amsfonts}
\usepackage{amsmath,bm}
\usepackage{amssymb}
\usepackage{amstext}
\usepackage{amsfonts}
\usepackage{graphicx,float}
\usepackage{mathptmx}
\usepackage{sectsty}
\usepackage{indentfirst}
\usepackage[svgnames]{xcolor}
\usepackage[font={scriptsize}]{caption}
\usepackage[colorlinks=True,linkcolor=Blue,citecolor=Blue,urlcolor=Blue,pdfstartview=FitH,bookmarks=False,pdfpagemode=UseNone]{hyperref}
\usepackage{epstopdf}
\usepackage{cite}
\usepackage{epsfig}
\usepackage{CJK}
\usepackage{booktabs}
\usepackage{multicol}
\usepackage{tikz}
\usetikzlibrary{patterns,calc,decorations.pathreplacing,decorations.markings}
\newcommand{\trace}{\mathop{\mathrm{tr}}}

\makeatletter
\def\@cite#1#2{\textsuperscript{[{#1\if@tempswa , #2\fi}]}}
\def\ExtendSymbol#1#2#3#4#5{\ext@arrow 0099{\arrowfill@#1#2#3}{#4}{#5}}
\def\RightExtendSymbol#1#2#3#4#5{\ext@arrow 0359{\arrowfill@#1#2#3}{#4}{#5}}
\def\LeftExtendSymbol#1#2#3#4#5{\ext@arrow 6095{\arrowfill@#1#2#3}{#4}{#5}}

\makeatother

\newlength{\halfpagewidth}
\divide\halfpagewidth by 2

\sectionfont{\fontsize{10.5}{10.5}\selectfont}

\begin{document}
\captionsetup[figure]{name={Fig.}}
	\begin{CJK*}{GB}{gbsn}  


\begin{center}
\bf Ultrasound attenuation in $s^\pm$-wave two-band  superconductors    
\end{center}

\footnotetext{\hspace*{-.45cm}\footnotesize $^\dag$ Corresponding author, E-mail: biaojin@ucas.ac.cn}

\begin{center}
\rm ZHAO Runzhong, JIN Biao$^{\dagger}$
\end{center}

\begin{center}
\begin{footnotesize} \sl
(School of Physical Sciences, University of Chinese Academy of Sciences Beijing 100049, China)
\end{footnotesize}
\end{center}

\begin{center}
\end{center}
\vspace{2mm}

\begin{center}
\begin{minipage}{15.5cm}
\begin{minipage}[t]{2.3cm}{\bf Abstract}\end{minipage}
The two-band $s^{\pm}$-wave state is currently considered to be the most promising candidate for newly discovered iron-based high-$T_c$ superconductors. In this work we study theoretically the ultrasound attenuation in $s^{\pm}$-wave two-band superconductors. The impurity effect is calculated within the $\mathcal{T}$-matrix approximation. In particular, our theory predict that, when the sizes of two order parameter are comparable, a Hebel-Slichter peak may show up in the ultrasound attenuation versus temperature curves. Our calculations also confirmed the presence of the resonant impurity scattering at low temperature, observed previously by other authors\cite{BCW09} in the calculation of the NMR relaxation rate $1/T_{1}$.
\end{minipage}
\end{center}

\begin{center}
\begin{minipage}{15.5cm}
\begin{minipage}[t]{2.3cm}{\bf Keywords}\end{minipage}
Ultrasound attenuation; Iron-based superconductors; $s^{\pm}$-wave pairing; $\mathcal{T}$-matrix approximation\\

\end{minipage}
\end{center}

\begin{multicols}{2}

\section*{Introduction}
The discovery of the high-$T_c$ superconductivity in iron-based compounds has promoted highly intensive  research activities in solid-state physics\cite{Stewart,WHD13,HKM11,CH15,HK15,H16,BS17}.
The maximum $T_c$ value of this iron-based family well exceeds that in MgB$_2$ and places them in the vicinity of cuprate superconductors. For this newly discovered  family of high-$T_c$ superconductors, the pairing symmetry of its superconducting gap is a key to understand the mechanism of superconductivity. Extensive experimental and theoretical work have been done to address this important issue for iron-based superconductors\cite{H16,J10,RQD15}.

From the  experimental perspective, the situation is unclear. The NMR Knight shift indicates the spin-singlet state\cite{Matano,Grafe,Ning}.
The transition temperature $T_{c}$ seems to be robust to impurity effects\cite{Kawabata}. While the penetration depth\cite{Hashimoto}, the specific heat\cite{Mu}, and ARPES \cite{Ding,Kondo} indicate almost isotropic two-gap system, the NMR relaxation rate $1/T_{1}$ shows no Hebel-Slichter coherent peak, nor the exponential decay at low temperature, but rather a power-like law, usually referred to as $T^3$ but in reality somewhere between $T^3$ and $T^{2.5}$, which suggest existence of line nodes\cite{Matano,Grafe,Nakai,Mukuda}.

Among many theoretical research studies, the state mostly accepted as a standard pairing state of the iron-based superconductors is the spin singlet $s$-wave state with sign-changing order parameter (OP), arising from  antiferromagnetic fluctuation\cite{MSJ08,Kuroki}. This state is currently referred to as the $s^{\pm}$-wave state. Sign-changing feature in the $s^{\pm}$-wave superconductor can produce some unique and distinct superconducting properties. In particular,  many authors have attempted to provide a theoretical explanation for the puzzling $1/T_{1}$ data mentioned above based on the $s^{\pm}$-wave state\cite{PDK08,BCW09,Chubukov}. It is demonstrated that due to the sign-changing feature of the $s^{\pm}$-wave superconductor, its  Hebel-Slichter coherent peak is substantially  suppressed intrinsically. The low temperature power-law-like behavior of $1/T_{1}$ is explained by the effect of  formation of  off-centered  resonant bound state inside the superconducting gap by impurity scattering in the unitary limit\cite{Bang}.

It is expected that the sign-changing feature of OP can also substantially affect the temperature dependence of the ultrasound attenuation $\alpha_{s}$ in $s^{\pm}$-wave superconductors since
the coherence factors for $1/T_{1}$ and $\alpha_{s}$ are related  closely similar to that in one-band case\cite{S64}. However, no  theoretical studies on ultrasound attenuation have been reported yet, to the best of our knowledge . It is the purpose of this work to narrow this gap.

\section{Formalism}
We start with the mean-field BCS Hamiltonian describing a two-band isotropic $s$-wave superconductor put forward by Suhl et al.\cite{SMW59}:
\begin{equation}\label{eq:Phai}
\begin{aligned}
		H &= \sum_{\mathbf{k},s}\varepsilon_{s\mathbf{k}}\left(c^\dagger_{s,\mathbf{k}\uparrow}c_{s,\mathbf{k}\uparrow}+c^\dagger_{s,-\mathbf{k}\downarrow}c_{s,-\mathbf{k}\downarrow}\right)\\
		 &- \sum_{\mathbf{k},s}\left(\Delta_{s}c^\dagger_{s,\mathbf{k}\uparrow}c^\dagger_{s,-\mathbf{k}\downarrow} + \Delta_{s}c_{s,-\mathbf{k}\downarrow}c_{s,\mathbf{k}\uparrow}\right)
\end{aligned}
\end{equation}
where $s(=1,2)$ denotes the band index, $\varepsilon_{s\mathbf{k}}$ is the energy of noninteracting state with respect to the chemical potential,  and $\Delta_{s}$, assumed to be a real quantity, is the OP defined  by
\begin{equation}
\Delta_{s} = \sum_{\mathbf{k'},s'}V_{ss'}\langle c_{s',-\mathbf{k'}\downarrow}c_{s',\mathbf{k'}\uparrow}\rangle,
\end{equation}
with $V_{ss'}$ being the effective electron-electron interaction within ($s=s'$) or between ($s\ne s'$) the bands. The temperature dependence of $\Delta_{s}$ is determined  by the  coupled self-consistent gap equations
\begin{equation}
\begin{aligned}
\Delta_{s}
=& \pi \sum_{s'} V_{ss'}N_{s'}T\sum_{n}\dfrac{{\Delta}_{s'}}{\sqrt{{\omega}_{n}^2 + {\Delta}_{s'}^2}}
\end{aligned}\label{gap}
\end{equation}
for $s$=1 and 2. In the above,  $N_s$ is  density of states  of the $s$th band at  Fermi energy, $T$ denotes the temperature, and $\omega_{n}=\pi T(2n+1)$ is the Matsubara frequency.

Now we calculate the ultrasound attenuation coefficient $\alpha_s $ for our superconducting state using the standard formula\cite{S64,N11}:
\begin{equation}
	\alpha_s \propto \lim_{\omega\to 0}\sum_{\mathbf{q}}{\mathrm{Im}\ \Pi^R(\mathbf{q},\omega)}
\end{equation}
where $\Pi^R(\mathbf{q},\omega)$ is Fourier transform of the retarded charge density correlation function
\begin{equation}
	\Pi^R(\mathbf{q},t) = -i\theta(t)\langle \left[\hat{\rho}({\mathbf{q}},t),\hat{\rho}(-\mathbf{q},0)\right]\rangle,
\end{equation}
with
\begin{equation}
	\hat{\rho}(\mathbf{q}) = \sum_{\mathbf{k},\sigma}c^\dagger_{\mathbf{k}\sigma}c_{\mathbf{k}+\mathbf{q}\sigma}.
\end{equation}
In practice, the correlation function is conveniently calculated  in the Matsubara frequency domain,
\begin{equation}\label{eq:Prealpha}
\begin{aligned}
\Pi(\mathbf{q},i\nu_m) =& \sum_{\mathbf{k},s,s'}T\sum_{n} \\
\bigg({G}_{s}&(\mathbf{k},i\omega_n){G}_{s'}(\mathbf{k} + \mathbf{q},i\omega_n + i\nu_m) \\
+ {G}_{s}&(-\mathbf{k},-i\omega_n){G}_{s'}(-\mathbf{k} - \mathbf{q},-i\omega_n - i\nu_m)\bigg)
	\end{aligned}
\end{equation}
where $G_{s}$ and $F_{s}$ denote  the usual normal and anomalous single-particle Matsubara Green's functions.
The summation over Matsubara frequency can be carried out with the help of the spectral representation of Green's functions,
and the result is then analytically continued to the real frequency axis $i\nu_m\rightarrow\omega+i\delta$. We finally obtain
\begin{equation}
\begin{aligned}
	\dfrac{\alpha_s}{\alpha_n} =& \int_{-\infty}^{\infty} d\omega\left(-\dfrac{\partial f}{\partial \omega}\right)\\
	\Bigg[&\left(\sum_{s}\dfrac{N_s}{N_{tot}}\mathrm{Re}\ \dfrac{{\omega}}{\sqrt{{\omega}^2 - {\Delta}_s^2}}\right)^2\\
	&-\left(\sum_{s}\dfrac{N_s}{N_{tot}}\mathrm{Re}\ \dfrac{{\Delta}_s}{\sqrt{{\omega}^2 - {\Delta}_s^2}}\right)^2\Bigg]
\end{aligned}\label{att}
\end{equation}
where $\alpha_{n}$ stands for the attenuation coefficient corresponding to the normal state, $f(\omega)$ is the Fermi distribution function, and $N_{tot}=N_{1}+N_{2}$.

\section{Ultrasound attenuation in the clean limit}
It is realized that, even for repulsive  intraband and interband interactions ($V_{ij} < 0$) the above gap equations produce  the so-called $s^{\pm}$-wave solutions when the interband pairing interaction $V_{12}(=V_{21})$ is dominant over the intraband interaction $|V_{12}|> |V_{11}| , |V_{22}|$\cite{MSJ08,GM95}; in the $s^{\pm}$-wave state the  two OPs acquire opposite signs.
In this paper we will discuss  the $s^{\pm}$-wave state  focusing on the   $|V_{12}|\gg |V_{11}| = |V_{22}|$ case.
\begin{figure}[H]
\centering\includegraphics[scale=0.3]{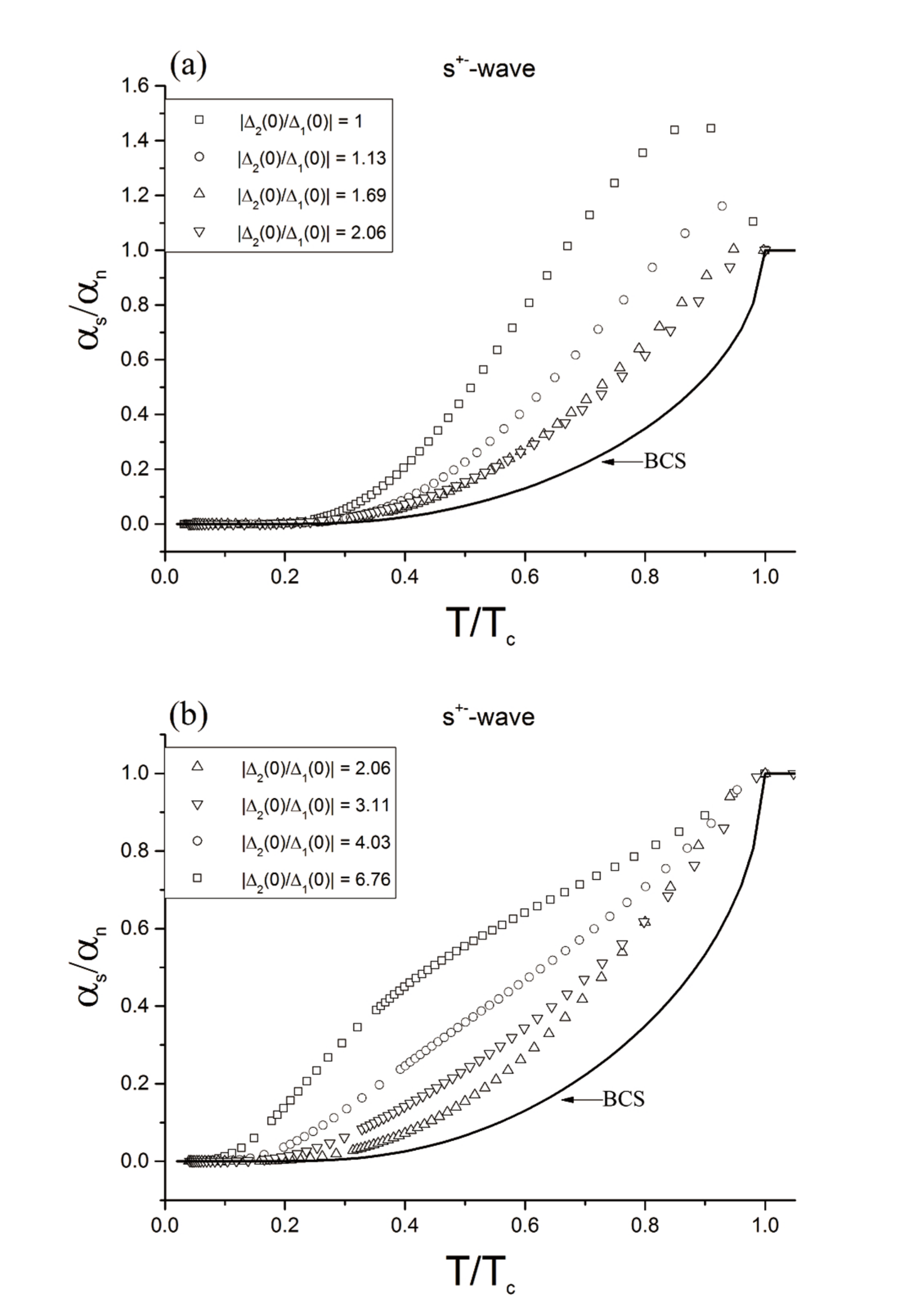}
\caption{Ultrasonic attenuation in clean $s\pm$-wave superconductor with gap ratio (a) $\le 2$, (b) $\ge 2$. One-band BCS result is shown with solid line for comparison.}
\label{fig:diffsym}
\end{figure}
Note that when $\Delta_{1}=\Delta_{2}$, $N_1=N_2$, Eq.(\ref{att}) reproduces the BCS result of one-band $s$-wave superconductor, $\alpha_s/\alpha_n =2 f(\Delta(T))$, which decreases exponentially at low temperatures. In the two-band case, however, the situation can be qualitatively different due to the presence of the crossing term $\propto\Delta_{1} \Delta_{2}$, especially when $\Delta_{1} \Delta_{2}<0$. It is known that such a  term (but with opposite sign) also shows up in the calculation of $1/T_{1}$ for the same $s^{\pm}$-model\cite{BCW09} leading to strong reduction of the Hebel-Slichter  peak. It is found that, in the case of the $s^{\pm}$-wave state $\alpha_s/\alpha_n$ exhibits a marked Hebel-Slichter peak like $1/T_{1}$, in sharp contrast to the one-band case.  We show in Fig. 1 the temperature dependence of  $\alpha_s/\alpha_n$ for various $|\Delta_{2}(0) /\Delta_{1}(0)|$, where $\Delta_{s}(0)$ is the zero-temperature value of the OP $\Delta_{s}(T)$. The solid line represents the BCS one-band result. As can be seen, the Hebel-Slichter peak reduces rapidly as the  ratio $|\Delta_{2}(0) /\Delta_{1}(0)|$ deviating from 1 (Fig. 1(a)). For $|\Delta_{2}(0) /\Delta_{1}(0)|>2$,  $\alpha_s/\alpha_n $ exhibits  typical behaviors of multiband superconductors; the fully gapped  exponential dependence is visible  only in the lowest temperature region (Fig. 1(b)).

\section{Impurity effects}

In this section,  we investigate impurity effects on ultrasound attenuation  employing  the self-consistent $\mathcal{T}$-matrix approximation\cite{HWE88,O04,MVH09}. It is well-known that, as a low impurity density expansion rather than a coupling constant expansion, the $\mathcal{T}$-matrix approximation can be used to describe impurity scattering continuously  from the Born (weak coupling) limit to the unitary (strong coupling) limit, hence can capture phenomena like the impurity resonance which does not possibly occur within the conventional Born approximation.

We consider an isotropic non-magnetic  impurity scattering\cite{O04} described by
\begin{equation}
	H_{\mathrm{imp}} = \sum_i\sum_{s,s',\mathbf{k},\mathbf{k'},\sigma}e^{i(\mathbf{k}-\mathbf{k'})\cdot\mathbf{R}_i}u_{ss'}c_{s\mathbf{k}\sigma}^\dagger c_{s'\mathbf{k'}\sigma}\label{Himp}
\end{equation}
where $u_{ss'}$ is single-impurity potential within ($s=s'$) or between ($s\ne s'$) the bands, and $\mathbf{R}_i$ is the position of the $i$-th impurity atom.
Using the Green's function approach one can take into account the impurity effects systematically within the scope of $\mathcal{T}$-matrix approximation (see Appendix). In the following calculations, we consider the $u_{ij}=u$ case.

Now due to the impurity scattering, the gap equations (Eq.(\ref{gap})) and the expression of the ultrasound attenuation coefficient (Eq.(\ref{att}))  are modified to
\begin{equation}
	\Delta_s = \pi\sum_{s'}V_{ss'}N_{s'}T\sum_{n}\dfrac{\tilde{\Delta}_{s'}}{\sqrt{\tilde{\omega}_{s'n}^2 + \tilde{\Delta}_{s'}^2}},
\end{equation}
and
\begin{equation}\label{eq:Alpha}
\begin{aligned}
	\dfrac{\alpha_s}{\alpha_n} =& \int_{-\infty}^\infty d\omega\left(-\dfrac{\partial f}{\partial \omega}\right)\\
	\Bigg[&\left(\sum_{s}\dfrac{N_s}{N_{tot}}\mathrm{Re}\ \dfrac{\tilde{\omega}_s}{\sqrt{\tilde{\omega}_s^2 - \tilde{\Delta}_s^2}}\right)^2\\
	&-\left(\sum_{s}\dfrac{N_s}{N_{tot}}\mathrm{Re}\ \dfrac{\tilde{\Delta}_s}{\sqrt{\tilde{\omega}_s^2 - \tilde{\Delta}_s^2}}\right)^2\Bigg].
\end{aligned}
\end{equation}

In the above, $\tilde{\omega}_{s}$ and $\tilde{\Delta}_s$ denote respectively the renormalized frequency and energy gap
\begin{align}
\tilde{\omega}_{s} &= \omega +i \Sigma_{s}^0(\omega),\\
\tilde{\Delta}_s &= \Delta_s + \Sigma_{s}^1(\omega),
\end{align}
with $\Sigma_{s}^0(\omega)$ and $\Sigma_{s}^1(\omega)$ being the impurity self-energy corrections to be determined in a self-consistent manner (see Appendix).

In discussing the impurity effects it is convenient to introduce two dimensionless parameters defined by $\Gamma = n_{\mathrm{imp}}/(\pi N_{tot})$ and $c=1/(\pi N_{tot}u)$ where  $ n_{\mathrm{imp}}$ denotes  the impurity concentration. Evidently, $\Gamma$  measures the impurity concentration and  $c$  the scattering strength. The Born limit and the unitary limit correspond to $c\gg1$ and $c\rightarrow 0$, respectively.

We study first the effect of impurity scattering in the Born limit ($c=10$). Plotted in Fig. 2(a) and Fig. 2(b) are the temperature variation of $\alpha_s/\alpha_n$ under different impurity concentration $\Gamma$ for $|\Delta_{2}(0) /\Delta_{1}(0)|=1$, and $|\Delta_{2}(0) /\Delta_{1}(0)|=1.8$, respectively. The one-band BCS result (the solid line) is also shown for the sake of comparison; within  the $\mathcal{T}$-matrix approximation, the Anderson's theorem \cite{Anderson} is valid for one-band $s$-wave state leading to  $\alpha_s/\alpha_n$ unaffected by non-magnetic impurity scattering (see Appendix).  Clearly, $\alpha_s/\alpha_n$ is rather robust for non-magnetic impurity scattering for $s^{\pm}$-wave superconductors in the Born limit .

\begin{figure}[H]
\centering
\includegraphics[scale=0.3]{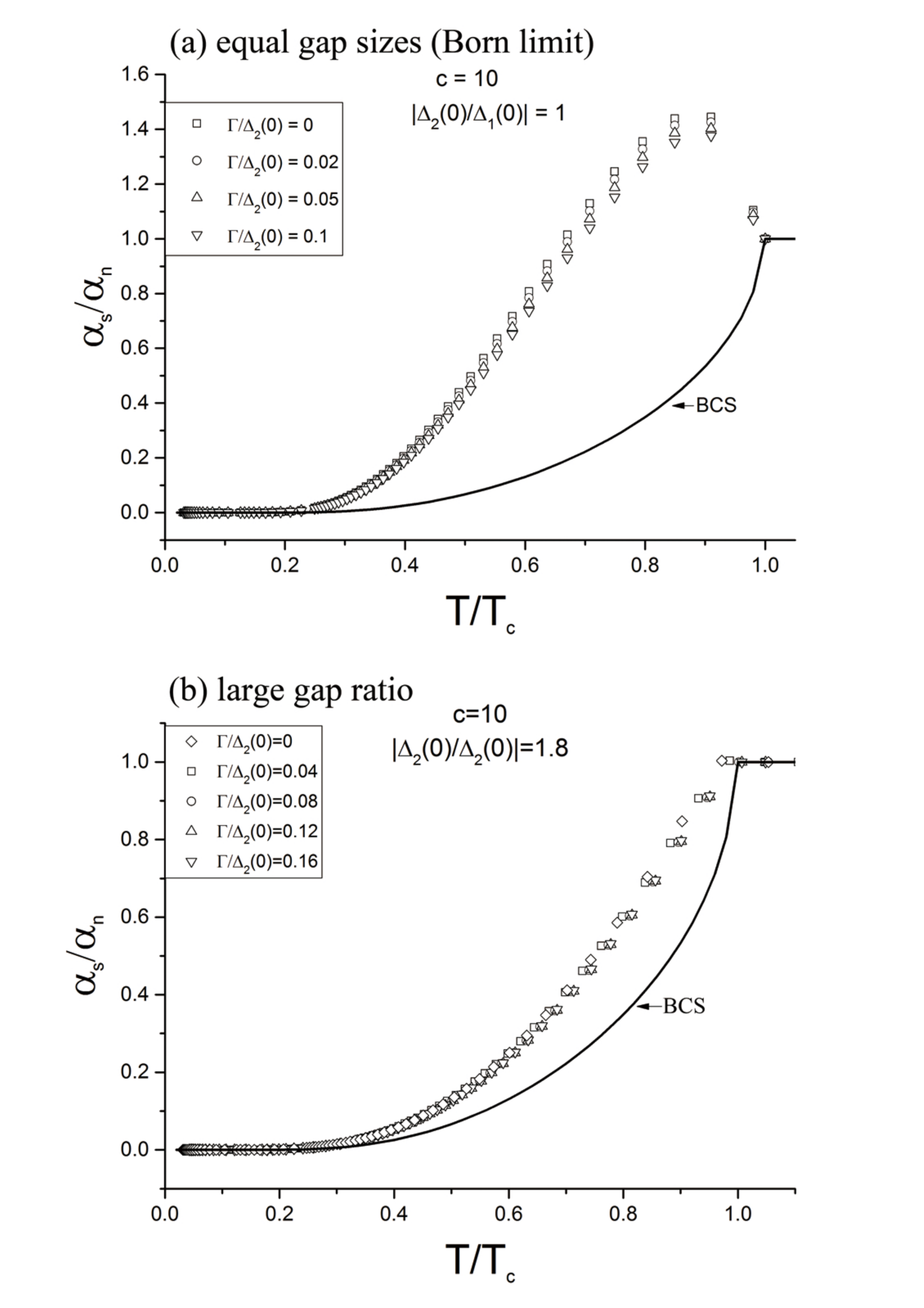}\\
\caption{Ultrasound attenuation in the Born limit with (a) equal gap sizes and (b) moderate gap ratio. Equal strength of intra- and inter-band scattering potential assumed. One-band BCS result is shown with solid line for comparison.}
\label{fig:lowtemp}
\end{figure}

Next let us turn to the case of unitary limit (c=0). Displayed in Fig.3 (a) and Fig. 3(b) are the temperature variation of $\alpha_s/\alpha_n$ under different impurity concentration $\Gamma$ for $|\Delta_{2}(0) /\Delta_{1}(0)|=1$, and $|\Delta_{2}(0) /\Delta_{1}(0)|=1.8$, respectively.
In the $|\Delta_{2}(0) /\Delta_{1}(0)|=1$ case, the $\alpha_s/\alpha_n$  as a whole, and the Hebel-Slichter coherence peak in particular, are suppressed with the increment of impurity concentration $\Gamma$ (Fig. 3(a)).  In the $|\Delta_{2}(0) /\Delta_{1}(0)|=1.8$ case, however, the temperature dependence of  $\alpha_s/\alpha_n$ is characterized by the absence of the Hebel-Slichter coherence peak, and the strong deviation from the fully gapped one-band behaviour at low temperatures due to the resonant  impurity scattering \cite{Bang}. Finally, we also have studied the cases with $|\Delta_{2}(0) /\Delta_{1}(0)|>4$, and found that  $\alpha_s/\alpha_n$ is almost unaffected by impurity scattering both in the Born and unitary limits.
\begin{figure}[H]
\centering
\includegraphics[scale=0.3]{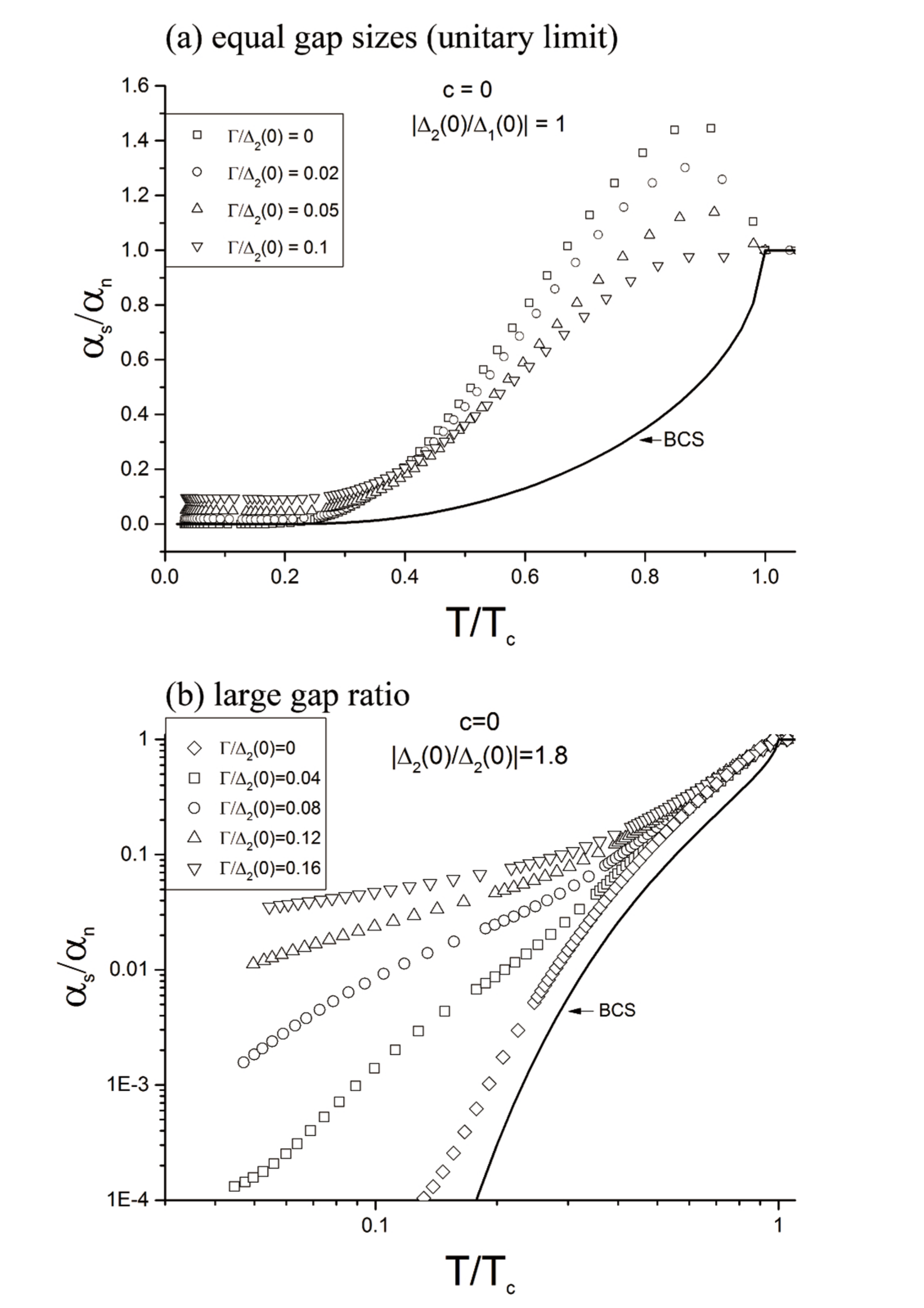}\\
\caption{Ultrasound attenuation in the unitary limit with (a) equal gap sizes and (b) moderate gap ratio. Equal strength of intra- and inter-band scattering potential assumed. One-band BCS result is shown with solid line for comparison.}
\label{fig:diffu}
\end{figure}

\section{Conclusion}
In this work, we have investigated the temperature dependence of the ultrasound attenuation coefficient, $\alpha_s/\alpha_n$, in the $s^\pm$-wave two-band superconductors with or without impurity scattering.
In discussing the impurity effects we have used the existing self-consistent $T$-matrix approximation, and considered both the Born and unitary limits. The important observation of this work is that,  when the sizes of two order parameter are comparable, a Hebel-Slichter peak may emerge in the  $\alpha_s/\alpha_n-T$ curves, in strong contrast to one-band case. This coherence peak is robust for impurity scattering in the Born limit, but  suppressed moderately  with the impurity concentration in the unitary limit. Experimental investigation on this issue is highly desired.
Besides, when $|\Delta_{2}(0) /\Delta_{1}(0)|$ is large, $\alpha_s/\alpha_n$ exhibit non-exponential-decay behaviors at low temperature, in consistent with that observed for $1/T_{1}$ in the previous theoretical investigations.

\section*{Appendix: $\mathcal{T}$-matrix approximation}
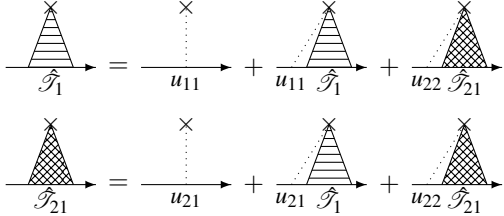
\begin{figure}[H]
\begin{center}
\begin{tikzpicture}
\draw[
	decoration={
        markings,mark=at position 1 with {\arrow{latex}}},postaction={decorate}]
	(-1.2,0) -- (0,0);
\fill[pattern=horizontal lines] (-0.9,0) -- (-0.6,0.8) -- (-0.3,0);
\draw (-0.9,0) -- (-0.6,0.8);
\draw (-0.3,0) -- (-0.6,0.8);
\draw node at (-0.6,0.8) {$\times$};

\draw node at (0.3,0) {$=$};
\draw[
	decoration={
        markings,mark=at position 1 with {\arrow{latex}}},postaction={decorate}]
	(0.6,0) -- (1.8,0);
\draw node at (1.2,0.8) {$\times$};
\draw[dotted] (1.2,0) -- (1.2,0.8);

\draw node at (2.1,0) {$+$};
\draw[
	decoration={
        markings,mark=at position 1 with {\arrow{latex}}},postaction={decorate}]
	(2.4,0) -- (3.6,0);
\fill[pattern=horizontal lines] (2.8,0) -- (3.1,0.8) -- (3.4,0);
\draw (2.8,0) -- (3.1,0.8);
\draw (3.4,0) -- (3.1,0.8);
\draw node at (3.1,0.8) {$\times$};
\draw[dotted] (2.6,0) -- (3.1,0.8);

\draw node at (3.9,0) {$+$};
\draw[
	decoration={
        markings,mark=at position 1 with {\arrow{latex}}},postaction={decorate}]
	(4.2,0) -- (5.4,0);
\fill[pattern=crosshatch] (4.6,0) -- (4.9,0.8) -- (5.2,0);
\draw (4.6,0) -- (4.9,0.8);
\draw (5.2,0) -- (4.9,0.8);
\draw node at (4.9,0.8) {$\times$};
\draw[dotted] (4.4,0) -- (4.9,0.8);

\draw node at (-0.6,-0.2) {\small{$\hat{\mathcal{T}}_{1}$}};
\draw node at (1.2,-0.2) {\small{$u_{11}$}};
\draw node at (2.6,-0.2) {\small{$u_{11}$}};
\draw node at (3.1,-0.2) {\small{$\hat{\mathcal{T}}_{1}$}};
\draw node at (4.4,-0.2) {\small{$u_{22}$}};
\draw node at (4.9,-0.2) {\small{$\hat{\mathcal{T}}_{21}$}};

\end{tikzpicture}
\begin{tikzpicture}
\draw[
	decoration={
        markings,mark=at position 1 with {\arrow{latex}}},postaction={decorate}]
	(-1.2,0) -- (0,0);
\fill[pattern=crosshatch] (-0.9,0) -- (-0.6,0.8) -- (-0.3,0);
\draw (-0.9,0) -- (-0.6,0.8);
\draw (-0.3,0) -- (-0.6,0.8);
\draw node at (-0.6,0.8) {$\times$};

\draw node at (0.3,0) {$=$};
\draw[
	decoration={
        markings,mark=at position 1 with {\arrow{latex}}},postaction={decorate}]
	(0.6,0) -- (1.8,0);
\draw node at (1.2,0.8) {$\times$};
\draw[dotted] (1.2,0) -- (1.2,0.8);

\draw node at (2.1,0) {$+$};
\draw[
	decoration={
        markings,mark=at position 1 with {\arrow{latex}}},postaction={decorate}]
	(2.4,0) -- (3.6,0);
\fill[pattern=horizontal lines] (2.8,0) -- (3.1,0.8) -- (3.4,0);
\draw (2.8,0) -- (3.1,0.8);
\draw (3.4,0) -- (3.1,0.8);
\draw node at (3.1,0.8) {$\times$};
\draw[dotted] (2.6,0) -- (3.1,0.8);

\draw node at (3.9,0) {$+$};
\draw[
	decoration={
        markings,mark=at position 1 with {\arrow{latex}}},postaction={decorate}]
	(4.2,0) -- (5.4,0);
\fill[pattern=crosshatch] (4.6,0) -- (4.9,0.8) -- (5.2,0);
\draw (4.6,0) -- (4.9,0.8);
\draw (5.2,0) -- (4.9,0.8);
\draw node at (4.9,0.8) {$\times$};
\draw[dotted] (4.4,0) -- (4.9,0.8);

\draw node at (-0.6,-0.2) {\small{$\hat{\mathcal{T}}_{21}$}};
\draw node at (1.2,-0.2) {\small{$u_{21}$}};
\draw node at (2.6,-0.2) {\small{$u_{21}$}};
\draw node at (3.1,-0.2) {\small{$\hat{\mathcal{T}}_{1}$}};
\draw node at (4.4,-0.2) {\small{$u_{22}$}};
\draw node at (4.9,-0.2) {\small{$\hat{\mathcal{T}}_{21}$}};

\end{tikzpicture}
\end{center}
\caption{Self-energy correction to Green's function in band $1$ within the $\mathcal{T}$-matrix approximation. The cross ($\times$) denotes the impurity site, and the solid line is impurity averiged Green's function.}
\label{fig:tm}
\end{figure}

For a two-band $s$-wave superconductor subjected to impurity scattering described
 by Eq.(\ref{Himp}), the impurity-averaged total single-particle Matsubara Green's function
can be written as
\begin{equation}
	\hat{\tilde{G}}_s^{-1}(\mathbf{k},i\omega_n) = \hat{G}_s^{-1}(\mathbf{k},i\omega_n) - \hat{\Sigma}_{s}(i\omega_n),
\end{equation}
where $\hat{G}_s$ is the Nambu Green's function in the absence of impurity scattering
\begin{equation}
	\hat{G}_s(\mathbf{k},i\omega_n) = -\dfrac{i{\omega}_{sn}\hat{\tau}^0 + \varepsilon_{s\mathbf{k}}\hat{\tau}^3+ {\Delta}_s\hat{\tau}^1}{{\omega}_{sn}^2 + \varepsilon_{s\mathbf{k}}^2 + {\Delta}_s^2}
\end{equation}
with $\hat{\tau}^i(i=0,1,2,3)$ the Pauli matrices, and  $\hat{\Sigma}_{s}(i\omega_n)$ the total self-energy matrix to be solved.
We can expand the self-energy in the form\cite{MVH09}:
\begin{equation}
	\hat{\Sigma}_s = \sum_i \Sigma_s^i\hat{\tau}^i
\end{equation}
with $\Sigma_s^2=\Sigma_s^3=0$ .
In our  $\mathcal{T}$-matrix approximation, where processes involving scattering from multiple impurity sites are ignored, the self energy is obtained by multiplying the single-impurity contribution by the impurity concentration, reflecting the multiple-scattering effect.\cite{M00}
\begin{equation}\label{eq:Sef}
	\Sigma_{s}^i(i\omega_n) = n_{\mathrm{imp}}\cdot \mathcal{T}_{s}^i(\omega_n)
\end{equation}
where $\mathcal{T}_{s}^i$ is the $\hat{\tau}^i$ component of the $\mathcal{T}$-matrix $\hat{\mathcal{T}_s}$.
In the two-band $s$-wave case, the equations for the $\mathcal{T}$-matrix for band $1$ are given by (see Fig.(\ref{fig:tm}) for a diagrammatic description)
\begin{equation}
\begin{aligned}
	\hat{\mathcal{T}}_1(i\omega_n) = \hat{u}_{11} &+ \sum_{\mathbf{k}}\hat{u}_{11}\hat{\tilde{G}}_1(\mathbf{k},i\omega_n)\hat{\mathcal{T}}_{1}(i\omega_n) \\
	&+ \sum_{\mathbf{k}}\hat{u}_{12}\hat{\tilde{G}}_2(\mathbf{k},i\omega_n)\hat{\mathcal{T}}_{21}(i\omega_n)\\
	\hat{\mathcal{T}}_{21}(i\omega_n) =  \hat{u}_{21}&+ \sum_{\mathbf{k}}\hat{u}_{21}\hat{\tilde{G}}_1(\mathbf{k},i\omega_n)\hat{\mathcal{T}}_{1}(i\omega_n) \\
	&+ \sum_{\mathbf{k}}\hat{u}_{22}\hat{\tilde{G}}_2(\mathbf{k},i\omega_n)\hat{\mathcal{T}}_{21}(i\omega_n).
\end{aligned}
\end{equation}
Combining these  equations above, one can obtain
\begin{equation}\label{eq:Tmat}
\begin{aligned}
	\mathcal{T}_{1}^0 =& -\dfrac{i}{D}\bigg[u_{11}^2g_{1,0}+u_{12}^2g_{2,0} \\
	&+ g_{1,0}(u_{12}^2- u_{11}u_{22})^2(g_{2,0}^2 + g_{2,1}^2)\bigg] \\
	\mathcal{T}_{1}^1 =&\dfrac{1}{D}\bigg[u_{11}^2g_{1,1}+u_{12}^2g_{2,1} \\
	&+ g_{1,1}(u_{12}^2 - u_{11}u_{22})^2(g_{2,0}^2 + g_{2,1}^2)\bigg]
	\end{aligned}
	\end{equation}
	where
	\begin{equation}
	\begin{aligned}
	D =& 1 + u_{11}^2(g_{1,0}^2 + g_{1,1}^2) + u_{22}^2(g_{2,0}^2 + g_{2,1}^2) \\
	&+ 2u_{12}^2 (g_{1,0}g_{2,0} + g_{1,1}g_{2,1}) \\
	&+ (u_{12}^2 - u_{11}u_{22})^2(g_{1,0}^2 + g_{1,1}^2)(g_{2,0}^2 + g_{2,1}^2)\\\\
	g_{s,0} =&\sum_k \dfrac{i}{2}\trace \hat{\tau}^0\hat{\tilde{G}}_{s}(k,i\omega_n) =\pi N_s\dfrac{\tilde{\omega}_{sn}}{\sqrt{\tilde{\omega}_{sn}^2 + \tilde{\Delta}_s^2}}\\
	g_{s,1} =&\sum_k \dfrac12\trace \hat{\tau}^1\hat{\tilde{G}}_{s}(k,i\omega_n)=  \pi N_s\dfrac{\tilde{\Delta}_s}{\sqrt{\tilde{\omega}_{sn}^2 + \tilde{\Delta}_s^2}}
\end{aligned}
\end{equation}
with the Matsubara frequency and gap functions renormalized as
\begin{equation}
\begin{aligned}
	\tilde{\omega}_{sn} &= \omega_n +i\Sigma_{s}^0(\omega_n)\\
	\tilde{\Delta}_s &= \Delta_s + \Sigma_{s}^1(\omega_n)
	\end{aligned}
	\end{equation}
The $\mathcal{T}$-matrices for the second band can be obtained by interchanging the band indices $1\leftrightarrow 2$.

In practice, one can solve Eqs. (\ref{eq:Sef}) and (\ref{eq:Tmat}) together with the coupled gap equations
\begin{equation}\label{eq:Gap}
	\Delta_s = \pi\sum_{s'} V_{ss'}N_{s'}T\sum_{n}\dfrac{\tilde{\Delta}_{s'}}{\sqrt{\tilde{\omega}_{s'n}^2 + \tilde{\Delta}_{s'}^2}}
\end{equation}
 self-consistently for $\mathcal{T}_{1}^0$, $\mathcal{T}_{1}^1$, $\Delta_{1}$, and $\Delta_{2}$, at different Matsubara frequencies, and  then obtain the retarded self-energies either via analytic continuation or by making use of the so-called Pad\'e approximants\cite{VS77,BCW09}.

\end{multicols}

\end{CJK*}
\end{document}